\documentclass[iop]{emulateapj}
\newcommand{\be}{\begin{equation}}
\newcommand{\ee}{\end{equation}}

\usepackage{color}

\usepackage{natbib}
\usepackage{enumerate}
\usepackage{graphicx}
\usepackage{xfrac}

\begin{document}
\title{Multiple Carbon monoxide Snow-lines in Disks Sculpted by Radial Drift}
\shorttitle{}
    \shortauthors{Cleeves}

 \author{L. Ilsedore Cleeves\altaffilmark{1,2}}

\altaffiltext{1}{Harvard-Smithsonian Center for Astrophysics, 60 Garden Street, Cambridge, MA 02138}
\altaffiltext{2}{Hubble Fellow}

\begin{abstract}
Observations of protoplanetary disks suggest that the gas and dust follow significantly different radial distributions. This finding can be theoretically explained by a combination of radial drift and gas drag of intermediate-sized dust grains. Using a simple parametric model to approximate the different distributions of the gas and dust components, we calculate and examine the impact of radial drift on the global dust temperature structure. We find that the removal of large grains beyond the ``truncation radius'' allows this region to become significantly warmer from reprocessed stellar radiation shining down from the disk upper layers, increasing the outer disk temperature by $\sim10-30\%$. This change is sufficient to raise the local temperature to a value exceeding the CO desorption temperature. These findings imply that the disk density structures induced by radial drift are able to create multiple CO snow-lines. The inner disk CO is in the gas phase, freezing out near the classical snow-line at $R\sim20-40$~AU. Moving outward, the CO sublimates once again beyond the truncation radius (80 AU in our models) and subsequently re-freezes out at sufficiently large stellar distances, beyond $R\gtrsim130-200$~AU. We find that thermal desorption of CO in the outer disk becomes competitive with external UV photodesorption and that this additional transition from solid state CO to the gas-phase has significant implications for the C/O ratio in the outer disk. 
\end{abstract}

\keywords{accretion, accretion disks --- astrochemistry ---  stars: pre-main sequence}

\section{Introduction}\label{sec:intro}

The gas and solids comprising protoplanetary disks follow different, yet linked evolutionary paths \citep[e.g.,][and references therein]{williams2011}. How (and when) this evolution occurs is essential in better understanding how planets form, and their resulting chemical compositions. While the physics affecting the gas distribution typically occurs on long timescales relative to the disk lifetime ($0.1-10$~Myr; e.g., through accretion, viscous dissipation, and photoevaporation), the dust is expected to evolve far more rapidly. For example, unhindered dust growth and subsequent settling will deplete the upper layers of disks in $\sim10^2-10^3$~years \citep[e.g.,][]{dullemond2005,brauer2008,birnstiel2010}, stopped only by turbulent mixing, collisional fragmentation \citep[e.g.,][]{dullemond2005}, or levitation by winds or magnetic fields \citep{turner2010}. Simultaneously, the slightly pressure-supported nature of the gas causes it to orbit at slightly sub-Keplerian speeds, causing the dust to experience a head-wind. Small grains are carried along with the gas through ``gas drag,'' while slightly larger particles lose substantial angular momentum to the gas and spiral inward through ``radial drift.'' 

While the idea of radial drift has long been recognized to be important \citep[e.g.,][]{whipple1972,weidenschilling1977}, \citet{birnstiel2014} recently demonstrated that gas drag and radial drift operating together result in a dust distribution that is sharply truncated at some small fraction of the full gas disk radius. For example, in their fiducial models, the gas extends beyond $\sim100$s of AU and the dust population (assuming an initial grain size of $r_g = 1\mu m$), evolves inward, resulting in a sharp outer truncation of the dust disk at $\sim80$~AU at late times. These models provided a natural explanation for the well-characterized observational feature that many disks exhibit significantly concentrated and sharply truncated millimeter dust emission compared to their distribution of gas (from CO) and/or small grains as traced by scattered light \citep[e.g.,][]{panic2009,andrews2012,degregorio2013}. 

The evolving redistribution of solids, both vertical and radial, will have significant implications for the disk thermal structure and the propagation of radiation through the disk. In the case of vertical settling, more settled disks tend to have warmer intermediate layers and colder midplanes \citep{dalessio2006}. Simultaneously, the removal of small grains from the disk atmosphere will decrease the local FUV opacity, increasing  the amount of gas exposed to stellar and external FUV photons, which enhances non-thermal desorption, drives photochemistry, and increases photodissociation \citep[e.g.,][]{fogel2011}.

The radial transport of solids is expected to alter the outer disk in similar ways, changing the outer disk thermal structure as well as its opacity to external UV radiation. In the following Letter, we explore the impact of radial drift on the global disk thermal structure using a simple two component dust model and examine the effects on the structure of the CO snow-line for low-mass stars. 

\section{Methods}\label{sec:physmod}
We have adapted the parametric density models of the dust and gas distributions from \citet{andrews2011}, which reflect the typical self-similarity solutions of \citet{lyndenbell1974}. The details of the implementation can be found in \citet{cleeves2013a}, where we just summarize the model here. The gas distribution is described by:
\begin{equation}\label{eq:sig}
\Sigma_{g} (R) = \Sigma_{c} \left( \frac{R}{R_c}  \right) ^{-\gamma} \exp{ \left[ -\left( \frac{R}{R_c} \right)^{2-\gamma} \right]},
\end{equation}
where $\Sigma_{c}$ is the surface density at the critical radius $R_c$. To simulate the effects of vertical settling, we adopt a two component dust model that consists of two separate MRN \citep{mrn1977} dust populations: a large grain population of $0.005\mu m$ to $1$~mm distributed over small scale heights, and a second mixed population of $a_{\rm max}=1$ and 10 micron-sized grains distributed with the same scale height as the gas \citep[see][for further details]{cleeves2013a}. We consider the percentage of mass in small grains a free parameter. 

The primary change in the present models from the previous treatment of \citet{cleeves2013a} is that we set the spatial distribution of the large grain population to have a different critical radius, $R_{c,{\rm{lg}}}$, from the gas and small grains, along with an additional truncation radius, $R_t$, to approximate the findings of the detailed models of \citet{birnstiel2014} with radial drift. We emphasize that this is a simple and flexible toy model and that more realistic, spatially dependent size distributions (including all of the important dust physics) should be explored in future work. For the large grains, we set $R_{c,{\rm{lg}}}=\sfrac{2}{3}~R_t$, and emphasize that the choice of $R_{c,{\rm{lg}}}$ does not strongly affect the results of this paper, where the large grains have a relative flat vertical distribution close to the midplane (see Section~\ref{sec:discussion}). The total gas mass is fixed over the models, $M_g=0.06$~M$_\odot$ between $R=0.1-600$~AU, with $\Sigma_c = 8.7$~g~cm$^{-2}$ and $R_c=100$~AU. The scale-height of the disk is 8~AU at 100~AU and varies as $h=H_{100}(r_{\rm AU}/100)^\psi$, with a flaring index of $\psi=0.3$.  We also assume that the total disk integrated gas-to-dust mass ratio is $f_g=100$ but varies with position.

We conserve the total dust mass, such that truncating the outer distribution of large grains at a location $R_t$ correspondingly increases the dust mass density in large grains. In reality, some amount of dust can (and likely will) be lost through direct accretion onto the star. The effect of this assumption is that the truncated large grain disk will be slightly cooler by a few Kelvin. The dust thermal structure is computed using the code TORUS \citep{harries2000,harries2004,kurosawa2004,pinte2009}, which utilizes the Lucy method \citep{lucy1999}. We have explored two low mass spectral types, a K4 type star (a future solar-type star) and an M1 type star, with effective temperatures of 4300~K and 3700~K and luminosities of 2.44 and 0.87~L$_\odot$, respectively. The corresponding masses and radii of the stars are 1.04~M$_\odot$ and 0.46~M$_\odot$ and 2.8~R$_\odot$ and 2.13~R$_\odot$, based upon the \citet{siess2000} online evolutionary tracks at 1~Myr.  We do not include heating due to accretion in the present models; however, it will have the largest effect in the inner few AU \citep[e.g.,][]{dalessio1998,dullemond2007}. We summarize the full parameter space modeled in Table~\ref{tab:modpar}.
\begin{deluxetable}{ll}
\tablecolumns{2} 
\tablewidth{0pt}
\tablecaption{Model Parameters. \label{tab:modpar}}
\tabletypesize{\footnotesize}
\tablehead{Type&Values}
Stellar spectral type & K4 (4300~K), M1 (3700~K)\\
$R_t$, large grain radius & 80~AU, Full Disk \\
Small dust mass fraction & 1\%, 10\% \\
CO freeze-out temperature & 17, 22, 26~K 
\enddata
\end{deluxetable}

\section{Results}\label{sec:results}
\subsection{Effects on the CO Snow-line}
We examine the temperature structure of the disk to predict the CO freeze-out regions. In detail, the location of a given snow-line is set by the balance of thermal desorption matched to freeze-out on grains. This location depends upon the number of surface sites, their coverage by other ices, and the binding energy, $E_b$, of the species. The CO binding energy furthermore depends on the substrate composition \citep[e.g.,][]{sa1988,sa1990}. CO frozen onto CO ice has a relatively low value of $E_b=855$~K, corresponding to a desorption temperature of $T_{\rm des}\sim17$~K \citep{oberg2005}, supported by the estimated temperature at TW~Hya's snow-line \citep{qioberg2013}. CO on CO$_2$ has an intermediate value, $E_b=1110$~K, or about $T_{\rm des}\sim22$~K. Finally, CO frozen onto water ice has the highest binding energy, $E_b(\rm{CO-H_2O})\sim1320$~K ($T_{\rm des}\sim26$~K), which agrees with the freeze-out temperature determined for the HD~163296 CO snow-line \citep{qi2015}. We consider this range of temperatures ($17-26$~K) in estimating snow-line locations. 

\begin{figure*}[ht!]
\begin{centering}
\includegraphics[width=1.0\textwidth]{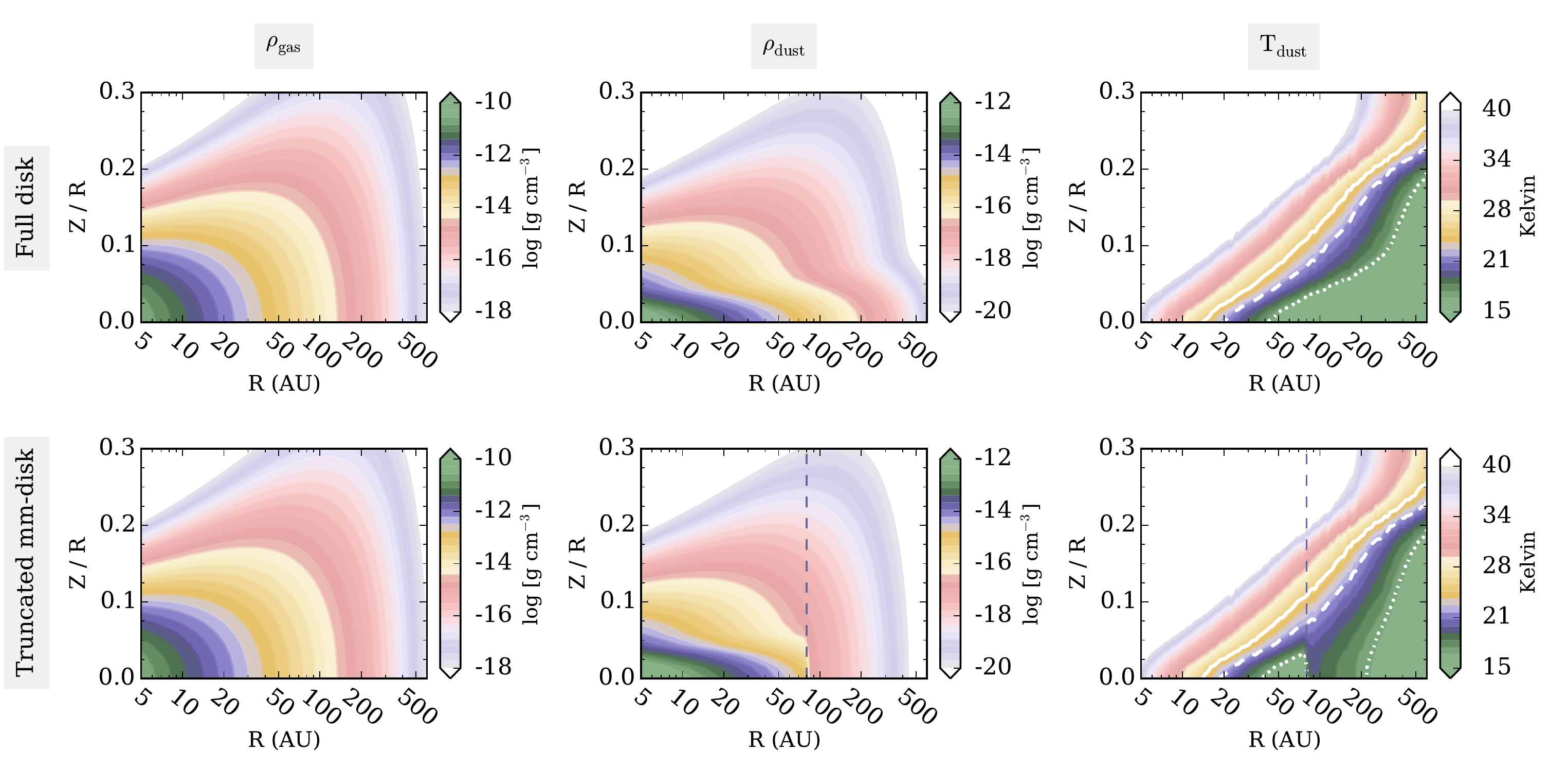}
\caption{Density and temperature for the $T_{\rm eff}=3700$~K model with 10\% dust mass in small (micron-sized) grains. The top row shows the model where large grains extend to the outer disk radius (Full disk) and the bottom row shows the truncated large grain disk case. The temperature plot color range is limited to highlight values near the CO sublimation temperature. The white contours indicate $T_{\rm dust}=17$~K (dotted), 22~K (dashed), and 26~K (solid). \label{fig:zoom}}
\end{centering}
\end{figure*}

Figure~\ref{fig:zoom} presents the density and thermal structure for a full disk and a model with an outer truncation at $R_t=80$~AU for just the large grains, imitating the effects of drift predicted in \citet{birnstiel2014}. Only the large grains are restricted to this smaller area, small grains are present throughout the extent of the gas distribution and contain 10\% of the total dust mass in this figure. We have highlighted the temperature isocontours at $T_{\rm dust}=17$~K, $22$~K, and $26$~K.  Compared to the full disk model, the truncated disk has a more complex temperature profile. The full disk has a monotonically decreasing temperature profile with distance from the star, causing there to be one midplane snow-line for a particular choice of $E_b$ (traced by the contour lines). By moving dust mass from the outer disk into the inner regions via radial drift, the opacity to the re-radiated stellar emission from the warm surface significantly drops. Consequently, the outer disk beyond the truncation radius is heated more than it would otherwise be without radial drift. 

One of the most important implications of the non-monotonic temperature profile in the drift case is that for reasonable assumptions, the outer disk can become warm enough to thermally desorb CO beyond the classical CO snow-line in the inner disk, typically observed in T~Tauri systems to occur near $R\sim20-40$~AU. As seen in Fig.~\ref{fig:zoom}, for $T_{\rm des}=17$~K, CO thermally desorbs outside of $R_t=80$~AU even at the midplane, and only starts to freeze out again at $\sim200$AU, where geometrical dilution of the stellar radiation field takes over.

\begin{figure*}
\begin{centering}
\includegraphics[width=0.8\textwidth]{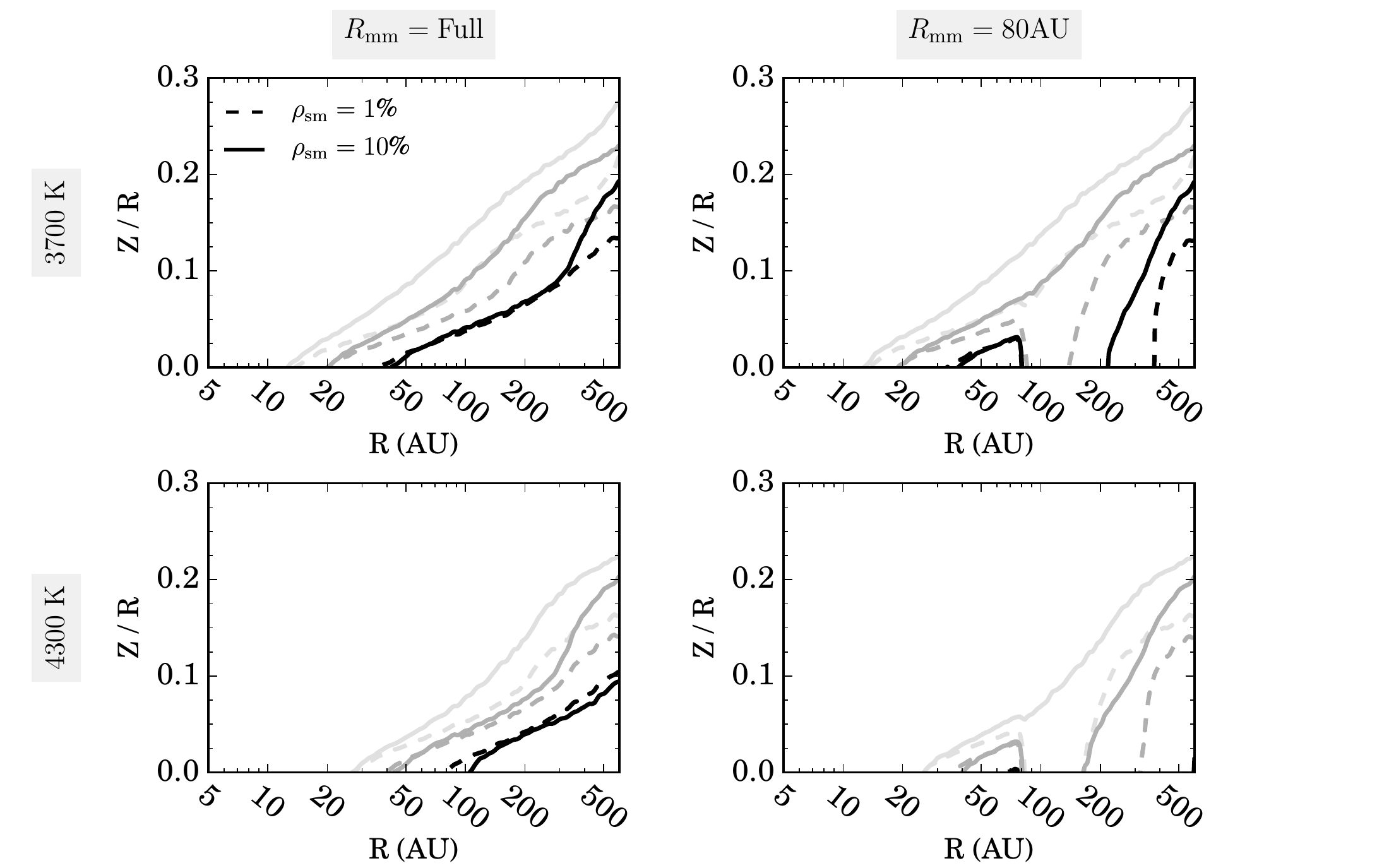}
\caption{Dust temperature isocontours at 26~K (light gray), 22~K (dark gray), and 17~K (black) for a stellar effective temperature of 3700~K (top row) and 4300~K (bottom row). The left column shows temperature contours for the full disk and the right column shows the truncated large grain case (i.e., with radial drift).  The line style indicates the fraction of mass in small grains where 10\% (solid) and 1\% (dashed). \label{fig:therm}}
\end{centering}
\end{figure*}

To test under what conditions this second CO evaporation front arises, we have varied the fraction of small grains and the central star's spectral type, Figure~\ref{fig:therm}. For the M-star (3700~K), decreasing the small grain mass fraction from 10\% to 1\% makes the intermediate layers of the disk warmer, and increases the extent of the outer CO sublimation region, spanning from 80 to 400~AU for $T_{\rm des}  = 17$~K, and 80 to 120~AU for $T_{\rm des}  = 22$~K.  For the highest value of the CO desorption temperature, $T_{\rm des}  = 26$~K, the CO desorption does not reach the outer disk midplane. 

Increasing the stellar mass to a young sun-like star ($T_{\rm eff}=4300$~K) also increases the size of the CO sublimation region, making the entire disk warmer than 17~K, except for a small region between $60-70$~AU. If we instead assume $T_d  = 22$~K, CO is frozen out from $40-80$~AU and beyond $\gtrsim170$~AU. Decreasing the small grain dust fraction increases this second CO freeze out region to beyond $\gtrsim300$~AU.

\subsection{Comparison to Photodesorption}
It is also of interest to compare thermal desorption to FUV photodesorption from the interstellar radiation field (ISRF), which is typically thought to dominate CO desorption in the outer disk. To test their relative importance, we calculate the flux of desorbed molecules due to thermal processes, i.e., Eq.~(3) of \citet{hollenbach2009}:
\begin{equation}
F_{\rm therm} = N_{s} R_{\rm therm} f_s,
\end{equation}
where $N_{s}=10^{15}$ is the number of adsorption sites per cm$^2$, $R_{\rm therm}$ is the per species rate of thermal desorption, Eq.~(2) of \citet{hollenbach2009}, and $f_s$ is the surface fraction covered by CO.  For the rate of thermal desorption, $R_{\rm therm}$, we take $E_b=855$~K, i.e., that of CO-CO.  We then calculate the flux of photodesorbed CO molecules from the ISRF with Eq.~(6) of \citet{hollenbach2009},
\begin{equation}
F_{\rm ISRF} = Y F_{\rm FUV} f_s,
\end{equation}
where $Y$ is the yield in molecules per photon and $F_{\rm FUV}$ is the photon flux incident on the grain. 

\begin{figure*} 
\begin{centering}
\includegraphics[width=0.89\textwidth]{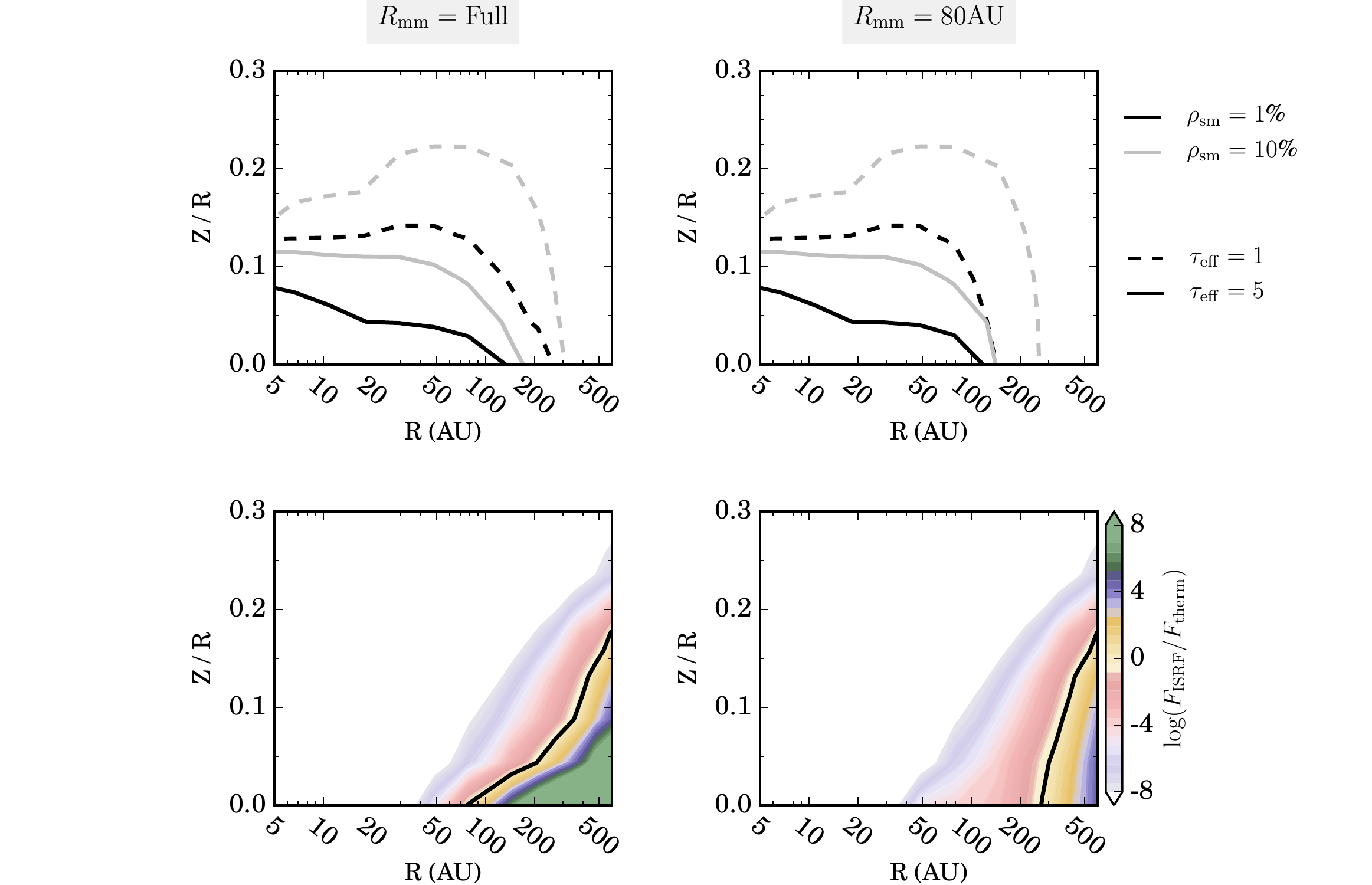}
\caption{Top row: Effective opacity ($\tau_{\rm eff}$) to external FUV radiation for the different small/large dust mass distributions over $4\pi$ steradians.  The left and right columns show the full model and drift-dominated model, respectively. The dashed and solid lines indicate $\tau_{\rm eff}=1$ and $\tau_{\rm eff}=5$. The line color corresponds to the different small grain dust fractions, gray (10\%) and black (1\%). Bottom row: For the $\rho_{\rm sm}=10\%$ case, the ratio of desorbed molecule flux from photodesorption vs. thermal desorption ($F_{\rm ISRF}/F_{\rm therm}$). The solid line indicates where the processes are equal. \label{fig:uvtau}}
\end{centering}
\end{figure*}

The mean ISRF is parameterized in units of $G_0$, $10^8$~photons~cm$^{-2}$~s$^{-1}$ between 6 and 13.6~eV \citep{habing}. We estimate the absorption opacity to each point in the disk from external radiation with the method of \citet{cleeves2013a} Appendix~A. This method samples the optical depth along rays that are uniformly spaced over $4\pi$ steradian and computes the ``effective'' optical depth ($\tau_{\rm eff}$) that a grain would see at a given location. This treatment better reflects the radiation field that the disk molecules experience compared to taking only the optical depth from the surface, which naturally underestimates the extinction. The $\tau_{\rm eff}$ curves are shown in Figure~\ref{fig:uvtau} (top) for both the truncated and full disks. At a given location, the optical depth of the outer disk is generally lower for the truncated disk with radial drift, as we would expect. 
Note that it does not move all the way inward to the truncation radius since there are still small grains out to the gas disk radius. 

We compute the local flux, $F_{\rm FUV} = G_0 \exp\left(-\tau_{\rm eff}\right)$, and adopt a value for the yield of photodesorbed CO of $Y=10^{-2}$~molecules~photon$^{-1}$, which is appropriate for the FUV photodesorption experiments of \citet{fayolle2011}. Figure~\ref{fig:uvtau} (bottom) compares the relative desorption fluxes from the ISRF to thermal processes ($F_{\rm ISRF}/F_{\rm therm}$) for the $\rho_{\rm sm}=10\%$ case. Large values correspond to efficiency in each process, i.e., regions of the disk where $F_{\rm ISRF}/F_{\rm therm}$ is large implies photodesorption dominates thermal desorption and vice versa. In all models, thermal desorption dominates the inner disk and transitions to photodesorption in the outer disk. We find that the transition is substantially further out for the drift-dominated model and that the transition occurs rapidly. In the case with drift, thermal desorption dominates ISRF photodesorption inside of $R<270$~AU and for the full disk the transition occurs at $\sim80$~AU. These results are non-obvious, as radial drift increases outer disk heating but also decreases the opacity to external FUV. We find similar results for 1\% small grain mass case, with values of 400~AU (drift) and 70~AU (full), respectively.

\section{Discussion}\label{sec:discussion}

Based on these results, we expect radial drift to alter the disk temperature profile and enhance thermal desorption of CO in the outer disk. To approximately estimate the enhancement in CO column density, we take the temperature profile and assign a CO abundance of $10^{-4}$ per H$_2$ for dust temperatures between 17~K and 200~K. We calculate the vertically integrated column density of CO versus radius and then compare the models with and without drift, and find the maximum enhancement in CO column is $\sim2-2.5$ times higher outside of the large grain radius. Recently, observations of IM~Lup have shown multiple rings of DCO$^+$, a direct product of CO-driven chemistry \citep{oberg2015im}. Furthermore, \citet{tang2015} report multiple CO ring structures in both $^{12}$CO and $^{13}$CO in GG~Tau. Radial temperature variations increasing CO sublimation in the outer disk may be contributing to these structures, though they should be modeled in detail to test non-thermal vs. thermal desorption scenarios. 

The extra CO sublimation changes the C/O ratio in the gas and ice in the outer disk. Similar to the calculations of  \citet{oberg2011}, we can estimate the C/O ratio in solids at the midplane. In a disk with well mixed gas and dust and a single CO snow-line, the C/O ratio in ices beyond the CO snow-line should be near the solar value because the main volatile sources of carbon and oxygen are fully frozen-out. The ``re-sublimation'' of CO beyond the truncation radius causes the C/O ratio in the gas to return to unity in this radial zone between the CO freeze-out regions, and correspondingly, the C/O ratio in the ice drops to sub-solar values, $\sim0.3$. 

\begin{figure*}[t]
\begin{centering}
\includegraphics[width=1.0\textwidth]{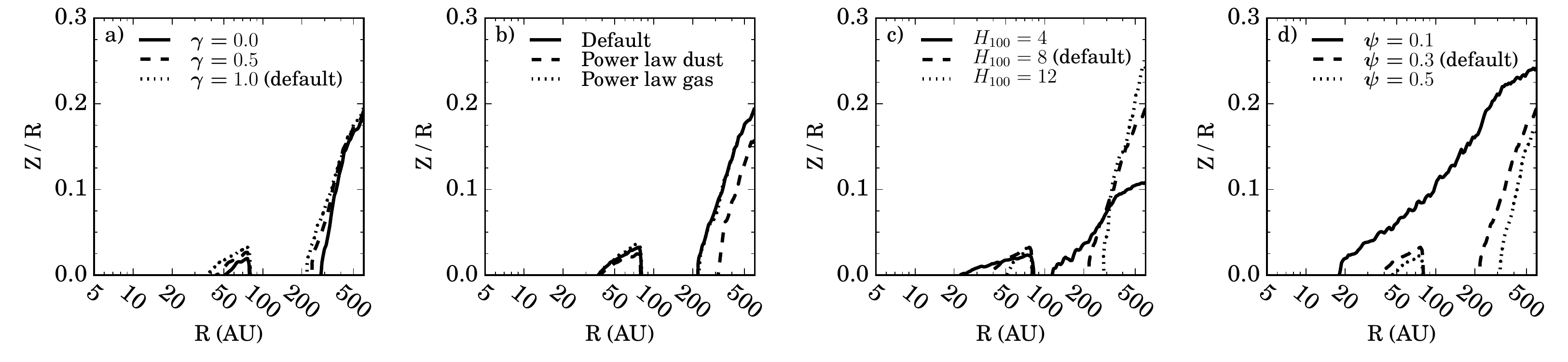}
\caption{Dependence on model parameters shown in the 17~K isotemperature contours for the 3700~K star with 10\% small dust mass. From left to right, {\em i.} disk temperature dependence on the surface density exponent $\gamma$ in Eq.~(\ref{eq:sig}), {\em ii.} mass surface density profile for a pure power law in large dust grains and in gas (normalized to contain the same disk mass), {\em iii.} disk scale height in AU at 100~AU ($H_{100}$), and {\em iv}. the flaring index $\psi$, see Section~\ref{sec:physmod}. \label{fig:paramtest}}
\end{centering}
\end{figure*}

We also explore how our results depend on the particular choices of model parameters in Figure~\ref{fig:paramtest}. We vary the distribution of mass ($\gamma$ in Eq.~\ref{eq:sig}); pure power law models in gas and dust normalized to the same disk mass; the gas scale height; and the flaring index, $\psi$, where larger values are more flared. We find the general structure of gas-phase CO in the inner disk, freeze-out in the large grain disk, re-sublimation beyond the truncation radius, and freeze-out again at large radii are robust features of the models, though in detail the snow-line locations will in change with the particular choice of parameters. The second snow-line disappears for the very flat disk case, $\psi=0.1$, where the disk is very dense and cold. 

These results apply to full disk models; however, there have been many recent observations of disks with large inner cavities in the dust \citep[e.g.,][]{casassus2013a,zhang2014,vandermarel2013,perez2014}. It has been previously seen that removing the inner disk does lead to a warmer disk near the inner edge \citep{cleeves2011}. In combination with the outer disk heating, in principle the inner ``classical'' snow-line could disappear inside of the truncation radius, leaving only the outer disk snow-line at $\sim100$s of AU. These effects may lead to the appearance of more distant snow-lines than the star might otherwise support for a full, non-transitional disk.

We have also calculated a temperature structure model for a Herbig-type disk with a $T_{\rm{eff}}=9333$~K star and twice the gas mass (1\% small grain model). In our models with drift, CO is present in the gas-phase everywhere. Nonetheless, there is still a temperature increase outside of the 80~AU truncation region, elevating the disk temperature from $T_d=30$ to over 40~K. In this case, other molecules with higher desorption temperatures near 40~K, such as HCN, may exhibit rings at the truncation radius.

Finally, after the discovery of ringed-structure in the HL~Tau protostellar disk \citep{alma2015}, there have been recent explorations matching the presence of rings to efficient dust-growth near prominent snow-lines \citep{zhang2015} or efficient ice-sintering (dust-growth suppression) near snow-lines \citep{okuzumi2015}. CO is a major constituent of the C and O reservoir, and in principle the decreased ice-content between the two CO-freeze out regions due to drift effects may regenerate small dust more efficiently by fragmenting collisions. However, it should be noted that these grains can still be water-ice coated and should still be generally ``sticky.'' A more detailed exploration of dust growth in the context of multiple snow-line structures is beyond the scope of this letter, but should be explored in future work.

\section{Summary}\label{sec:summary}
Using simple parameterized disk models, we explore how a more compact large grain distribution relative to the distributions of gas and small grains alters the temperature structure of the outer disk. Such distributions are a natural outcome of a combination of radial drift and gas drag \citet{birnstiel2014} and are a commonly observed phenomenon. We find that the removal of dust mass from the outer disk causes this region to become sufficiently warm to support direct thermal desorption of CO. In our models with a large-grain truncation radius at 80~AU \citep[based on the findings of][]{birnstiel2014}, CO can thermally desorb from this radius out to $R\sim100-400$~AU depending on properties of the star and disk. Moreover, thermal desorption of CO becomes competitive with external UV photodesorption from the interstellar radiation field. These results have important implications for the gas versus solid state C/O ratios of protoplanetary disks, where CO is a major carbon and oxygen reservoir.

\acknowledgements{Acknowledgements: The author is grateful to the anonymous referee whose helpful comments improved this paper. The author also thanks Karin {\"O}berg, David Wilner, Ryan Loomis, and Andrea Isella for helpful discussions. LIC acknowledges the support of NASA through Hubble Fellowship grant HST-HF2-51356.001-A awarded by the Space Telescope Science Institute, which is operated by the Association of Universities for Research in Astronomy, Inc., for NASA, under contract NAS 5-26555. 
}

\end{document}